\documentclass[amsmath,amssymb, aip, jap, twocolumn, showkeys, reprint]{revtex4-2} 

\usepackage{amssymb}
\usepackage{amsmath}
\usepackage{color}
\usepackage{graphicx}

\usepackage[version=3]{mhchem} % Formula subscripts using \ce{}
\usepackage[dvipsnames]{xcolor}
\usepackage{textgreek}
\usepackage[detect-all]{siunitx}

\usepackage{hyphenat}
\usepackage{pgf,tikz}
\usepackage[english,czech]{babel}

\usepackage{upgreek}

\usepackage{geometry}
\newcommand{\add}[1]{#1}

\newcounter{exception}
\newcommand{\excep}[1]{\refstepcounter{exception}\label{#1}\ref{#1}}

\usepackage{dcolumn}
\usepackage{bm}

\begin{document}

%\preprint{AIP/123-QED}

\title{Essential Principles and Practices in X-ray Photoelectron Spectroscopy}

\selectlanguage{Czech}
\author{Jan Čechal}
\selectlanguage{english}
 \email{cechal@fme.vutbr.cz}
\affiliation{ 
CEITEC and Institute of Physical Engineering, Brno University of Technology, Brno, Czech Republic.
}%

\date{\today}

\begin{abstract}
\noindent X-ray photoelectron spectroscopy (XPS) is a widely used technique for chemical analysis of solid surfaces, sensitive to the chemical environments of atoms via core-level binding energy shifts. While modern instruments allow  experimental data to be acquired with ease, their evaluation and interpretation remain challenging for newcomers to the field, as a profound knowledge of the method is required for correct analysis. Here we present a concise yet comprehensive \add{overview} of the fundamental principles and methodologies of XPS, covering photoemission processes, chemical shifts, charge referencing, peak fitting, and quantification strategies. This overview aims to bridge the gap between data collection and reliable analysis, providing essential knowledge for correct interpretation. By clarifying key concepts and common practices, this work supports improved accuracy in surface chemical characterization using XPS.
\end{abstract}

%Use showkeys class option if keyword
                              %display desired
\maketitle

%\tableofcontents

\section{Introduction}

\noindent X-ray photoelectron spectroscopy (XPS) is a method for chemical analysis of near-surface layers of materials. Its primary advantage is its sensitivity to the chemical environments of probed atoms as revealed by shifts in binding energies (BE) of core levels. In recent years, the widespread adoption of this method has been accompanied by a high rate of incorrect analyses and interpretations, reaching as much as 60\,\%, \cite{major2020} which has been identified as a pre-crisis situation. \cite{Major2023} Improving education and raising awareness are essential to prevent a full-blown crisis in the field. \cite{Major2023} While many resources are available –- including the well-known “red” \cite{Briggs1990} and “blue” \cite{Briggs2003} books on XPS, reviews, \cite{Greczynski2022,Greczynski2023,Chambers2024} series of tutorial papers, \cite{Baer2021} and a guideline to avoid common errors \cite{Pinder2024} –- these are often lengthy and not easily accessible for quick reference. In this work, we provide a brief yet comprehensive introduction to XPS, serving as a signpost to guide readers to the source of knowledge they may need. To maintain conciseness, we do not discuss exceptions to general rules that are valid in more than 90\,\% of cases; such exceptions are marked with “(!\,23)”, where the number refers to a note in Appendix A.

\vspace{-3mm}

\section{Photoelectron Spectrum}

\noindent \textbf{Process of photoemission.} In XPS, a vacuum-compatible sample is irradiated by monochromatic Al K$\alpha$ X-rays (!\,\excep{monochrom}) of energy $E_\mathrm{Al\ K\alpha}=1486.6$\,eV (Fig.\,1A). X-rays penetrate a few micrometers into the sample (Fig.\,1C), where they can be absorbed by an atom, which subsequently emits a photoelectron from a core level (CL) with a binding energy of $E_\mathrm{CL}$ (Fig.\,1F). The kinetic energy of the \add{emitted} photoelectron is $E_\mathrm{kin}^\ast$. The energy distribution of the measured electron kinetic energies $E_\mathrm{kin}$ is referred to as a photoelectron spectrum (Fig.\,1H). We treat photoemission, photoelectron transport through the solid, and escape into vacuum as independent events. The spectrum also contains contributions from other processes, such as X-ray-induced Auger electrons and inelastically scattered electrons, which form a background. Due to inelastic scattering during transport through the solid, only electrons emitted within the topmost few nanometers of the sample can escape \add{without losing their characteristic energy} (Fig.\,1C). \cite{Greczynski2022,Powell2020}

\begin{figure*}
\includegraphics[scale=0.85]{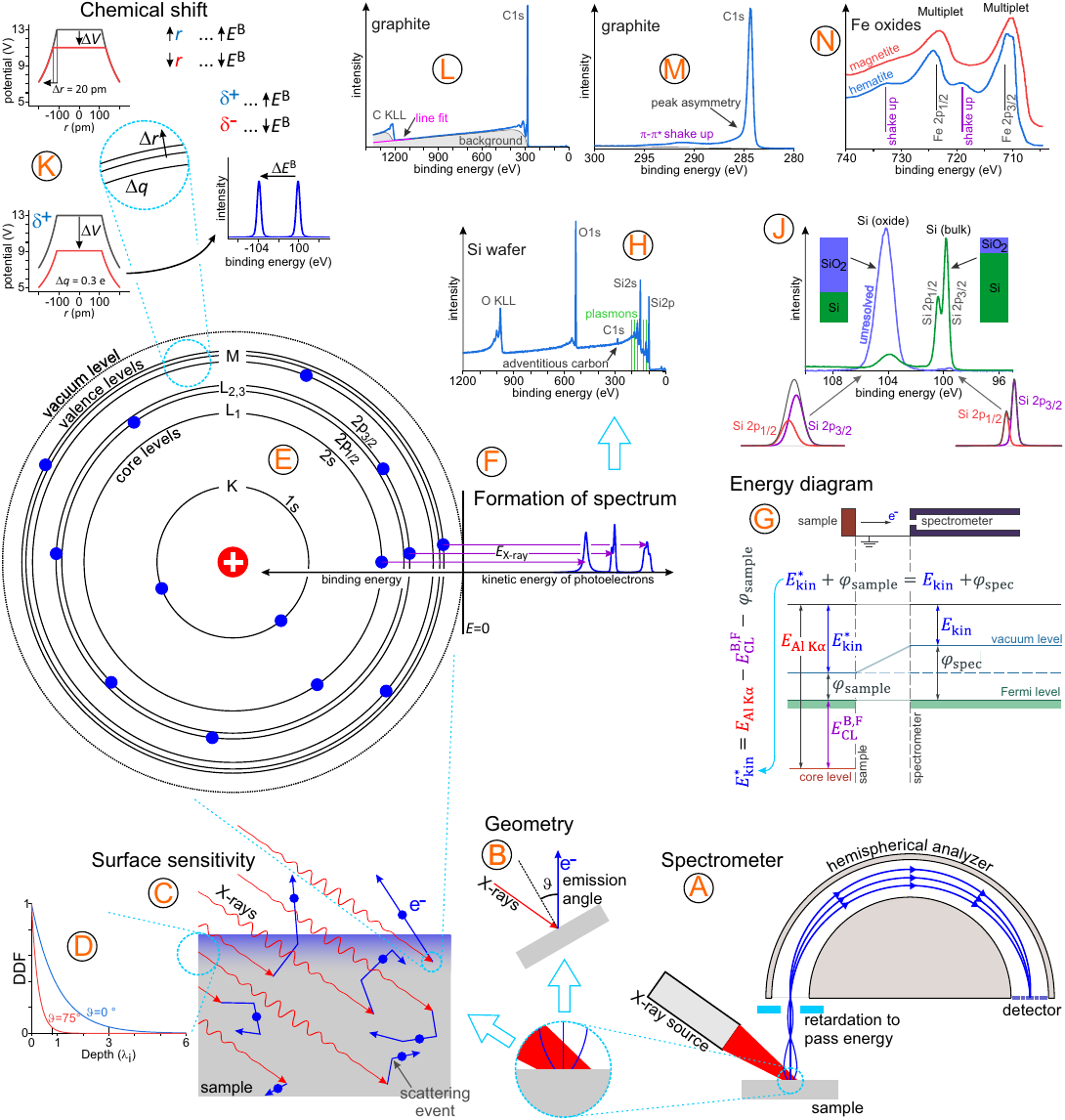}
\caption{\label{fig} \add{Overview summary of XPS processes and measurements}. 
(A) Typical experimental setup. 
(B) Definition of an emission angle. 
(C) X-rays penetrate $\upmu$m into the sample, whereas the electrons can escape only from a few nm according to the depth distribution function (DDF) given in (D) for two distinct emission angles defined in (B). 
(E) Schematics of the probed atom with marking of inner shells in X-ray and spectroscopic notations. 
(F) Formation of the spectrum by excitation of electrons with radiation of constant energy of  $E_\mathrm{X\text{-}ray}$. 
(G) Energy diagram defining energy conservation rules. Equation (1) is obtained by combining energy conservation for sample–spectrometer contact and photoemission events. 
(H) Photoelectron spectrum of Si wafer; vertical lines highlight plasmon loss peaks. 
(J) Detailed Si\,2p spectra of two distinct thicknesses of oxide layer on bulk Si. Both peaks are doublets; their fitting is detailed in the bottom part. 
(K) Charged shell model for chemical shifts. Changing either valence charge density by $\Delta q$ of the mean radius of the valence level $\langle r\rangle$ changes the inner potential within the shell, which changes the BE of electrons on core levels. The potential is given as a function of $\langle \frac{1}{r}\rangle$; inside the shell, the potential is constant. 
(L) Photoelectron spectrum of graphite showing with marked background and straight line fit of pre-peak intensity for C\,KLL.
(M) Detailed C\,1s spectrum of graphite showing its inherent asymmetry and shake-up satellite. 
(N) Detailed Fe\,2p spectra of pure Fe oxides (magnetite and hematite) showing broad peaks of complicated shape due to multiplet splitting. 
}
\end{figure*}

Energy conservation relates the measured kinetic energies of electrons to the BEs of core levels from which they were emitted (Fig.\,1G). BEs are referenced typically to the sample Fermi level $E_\mathrm{CL}^\mathrm{B,\,F}$ or more rarely to vacuum level $E_\mathrm{CL}^\mathrm{B,\,V}$; these energy levels are mutually \add{separated} by the sample work function $\varphi_\mathrm{sample}$: $E_\mathrm{CL}^\mathrm{B,\,V}=E_\mathrm{CL}^\mathrm{B,\,F}+\varphi_\mathrm{sample}$. For \textit{conductive} samples, a common Fermi level is established between the sample and the spectrometer by grounding both; then the energy conservation relation is 
\begin{equation}
E_\mathrm{kin} = E_\mathrm{Al\ K\alpha} - E_\mathrm{CL}^\mathrm{B,\,F} - \varphi_\mathrm{spectrometer},
\end{equation}
featuring a constant spectrometer work function $\varphi_\mathrm{spectrometer}$ (Fig.\,1G). \cite{Greczynski2024a,Chambers2024} Hence, for grounded conductive samples, the measured BEs are independent on $\varphi_\mathrm{sample}$, making the Fermi-level referencing a general approach for a variety of samples. Equation (1) is used for the conversion of the axis of kinetic energies to the axis of binding energies; it is a convention to plot the binding energy axis with BE increasing to the left. For \textit{non-conductive} (insulating) samples or physisorbed species, a common Fermi level is not established, and Equation (1) is no longer valid; it is replaced by
\begin{equation}
E_\mathrm{kin} = E_\mathrm{Al\ K\alpha} - E_\mathrm{CL}^\mathrm{B,\ F} - \varphi_\mathrm{sample}+V,   
\end{equation}
where $V$ is a potential \add{at the surface} due to, e.g., employed charge neutralization. \cite{Greczynski2024a} Insulating samples would charge positively due to emission of electrons; to prevent such charging, samples are supplied by low-energy electrons from the vacuum side (!\,\excep{charneut}), which is referred to as charge neutralization. \cite{Baer2020,Crist2024} We note that charge neutralization is not perfect, and artifacts can arise from differential charging \add{(!\,\excep{neutral})}.

\textbf{Core levels}. If an atom of a given element is present in the probed volume, all its accessible core-level peaks are observed with characteristic relative intensities; the sensitivity is $\sim\,$0.1–-1 atomic \%. \cite{Stevie2020} The BEs of core-levels, obtained from Equation (1), depend on the chemical \add{environment} of the atom, such as bonding to other atoms or its position within the crystal lattice. Peaks associated with core levels are labeled according to the quantum numbers of the level from which the electron was emitted (Fig.\,1E). For example, $2\mathrm{p}_{3/2}$ refers to emission from the level with principal quantum number $n = 2$, azimuthal quantum number $l = 1$ (labeled as s, p, d, and f for $l$= 0, 1, 2, and 3, respectively), and the total angular momentum number $j=3/2$, where $j$ is the sum of orbital and spin angular momenta ($l\pm\frac{1}{2}$). Levels with $l= 1$ (p), 2 (d), or 3 (f) are split by spin-orbit coupling into two components with different $j$ values (Fig.\,1J); these components have characteristic energy separations and intensity ratios (!\,\excep{photoem}) of 1:2, 2:3 and 3:4 for p, d, and f levels, respectively. \cite{Greczynski2022,Stevie2020} Both components represent the same chemical state.

\textbf{Chemical shifts.} Chemical (BE) shifts provide a direct way to characterize the chemical environment of probed atoms. The relationship between the chemical state and the measured BE position of the peak is not unique: \add{the same chemical environment should result in the same BE peak position, but} distinct chemical states can yield the same BE. We distinguish between initial-state effects, which occur before the photoemission event, and final-state effects, which occur in a response to photoemission. \cite{Bagus2023,Baer2020a} Both types of effects depend on the valence charge density and its distribution (!\,\excep{chemshift}). The initial-state effect can be illustrated (Fig.\,1K) by modeling the valence electrons as a thin spherical shell of mean radius $\langle r\rangle$ carrying charge $q$. \cite{Briggs1990, Bagus2023} The potential inside this shell is constant and proportional to $q\langle \frac{1}{r}\rangle$; this potential reduces the effective potential experienced by the core-level electrons, thereby affecting their binding energy. Increasing the valence charge density or decreasing the mean radius lowers the BE, whereas decreasing the charge density or increasing the radius raises the BE. Final-state effects involve the response of the atom to the formation of a core hole, primarily through screening by valence electrons. \cite{Bagus2023} 
\add{We note that no photoemission occurs from atoms that are absent (!\,\excep{Ovac}).}

\textbf{Charge referencing.} For \add{semiconducting, insulating, multiphase, or multilayer samples, in which charge can accumulate at interfaces,} establishing correct values of BE is challenging. \cite{Greczynski2024a} \add{For highly insulating specimens or those electrically isolated from ground}, Equation (2) must be used, and \add{even after charge neutralization}, the spectrum is shifted by an unknown value of $V$. A commonly used approach is to shift the spectrum so that the C\,1s peak of adventitious carbon (AdC, Fig.\,1H) appears at 284.8\,eV. \cite{Biesinger2022} 
\add{However, since the AdC aligns with the vacuum level \add{(not the Fermi level)}, \cite{Greczynski2024} Equation (1) does not apply to AdC even for grounded conductive samples, and the position $E_\mathrm{C\,1s}^\mathrm{B,\,F}$ of the C 1s peak is work-function dependent. \cite{Greczynski2024a,Cechal2026} As a result, the AdC C 1s peak position referenced to the Fermi level spans a 2.9\,eV interval.\cite{Greczynski2024a} A separate dataset from a multiuser facility shows that the AdC C 1s peak lies within (284.9$\pm$0.5)\,eV in 95\,\% of samples with an available secondary reference\cite{Biesinger2022}  (!\,\excep{C1senergy}). Hence, the correction using AdC needs to be applied with care to give satisfactory results. \cite{Biesinger2022,Greczynski2024a,Morgan2025} Often, it is better to use the known chemical state of another element as a reference.\cite{Baer2020}
In the case of semiconducting samples, band bending leads to a shift and/or a perturbation in the shape of all CL peaks. \cite{Chambers2024}}

\textbf{Peak fitting}. The primary goal in XPS is to identify chemical environments (e.g., bonding) of atoms in the sample from the position of their peaks, using reference values of chemical shifts. When multiple chemical environments are present, measured peaks may have complex shapes that are typically decomposed by modeling them with synthetic component peaks (Fig.\,1J); this procedure is referred to as peak fitting. \cite{Major2020a, Sherwood2019} Often, a spectrum is fitted with a set of peaks described by Voigt functions, which are convolutions of Lorentzian and Gaussian functions. \cite{Pinder2024,Jain2018,Chambers2024} The Lorentzian component is associated with the photoemission process, and its width is proportional to the core-hole lifetime. \cite{Greczynski2022, Krause1979,Chambers2024} The Gaussian component reflects instrumental broadening or broadening due to variations in the chemical environment, such as in amorphous oxides (Fig.\,1J). \cite{Greczynski2022,Chambers2024} For inherently asymmetric peaks, the Lorentzian is replaced by an asymmetric function like the Doniach--Šunjić function. \cite{Greczynski2022,Major2021} Spin-orbit-split components are modeled as doublet peaks (doublets) with given energy separations and intensity ratios (see above); both components represent a single chemical environment. It should be noted that not every spectrum requires fitting; in many cases, the necessary information can be obtained without it.

\section{Other Features in Photoelectron Spectrum} 

\noindent The photoelectron spectrum contains a wealth of information; experimentalists are encouraged to utilize the full spectrum, rather than just core-level peaks. 

\textbf{Shake-up satellite peaks} arise due to electron excitation accompanying the photoemission process, leading to a characteristic loss of energy. \cite{Briggs2003} The presence or absence of these satellites is indicative of specific chemical environments; \cite{Kocklauner2025} for example, the presence of a $\pi-\pi^\ast$ satellite (Fig.\,1M) is characteristic of aromatic systems. \cite{Gardella1986} The intensity of shake-up satellites should be included in quantification (!\,\excep{theosensf}). \cite{Brundle2020} A similar effect, involving excitation of electrons in the conduction band near the Fermi level, leads to the inherent asymmetry of peaks (Fig.\,1M) associated with many conductors. \cite{Greczynski2022} 

\textbf{Multiplet splitting.} Photoemission from certain atoms, particularly transition-metal atoms in paramagnetic states, can result in multiple peaks associated with a single chemical state due to coupling of the high angular momentum of the atom with the angular momentum of the created core hole. \cite{Bagus2023,Shirley1978, Chambers2024} For example, both hematite and magnetite (Fig.\,1N) should be represented with multiple (more than four) overlapping peaks, rather than a single component, to accurately reflect their multiplet splitting. \cite{Hughes2025, Chambers2024} \add{We note that even s levels can be split as in the case of Mn 3s. \cite{Chambers2024}}

\textbf{Auger peaks}. The relaxation of atoms left with a core hole after photoemission can proceed via the Auger process. A series of Auger peaks (Fig.\,1H,\,L) is a typical feature found in many XPS spectra. \cite{Briggs1990} Auger peaks provide additional information beyond that from core levels, especially when the kinetic energy of an Auger peak $E_\mathrm{AE}$ is combined with the binding energy of a core level $E_\mathrm{CL}^\mathrm{B,\,F}$ of the same element to form the modified Auger parameter $\alpha^\ast=E_\mathrm{CL}^\mathrm{B,\,F} + E_\mathrm{AE}$; the values of $\alpha^\ast$ enable chemical speciation in cases where chemical shifts alone are insufficient. \cite{Moretti1998,Baer2025} In some cases, the visual inspection of the Auger peak shape and its comparison with reference spectra can provide the necessary information about the chemical environment of the studied atoms.

\textbf{Extrinsic effects.} Additional features arise during electron transport through the solid. For example, inelastic scattering of photoelectrons gives rise to background signals (Fig.\,1L) \cite{Tougaard2021} while the excitation of plasmons leads to a series of loss peaks (Fig.\,1H) \cite{Stevie2020}. These plasmon-loss peaks are characterized by constant energy separations and decreasing intensities with increasing binding energy.

\section{Photoelectron Transport through a Solid}

\noindent \textbf{Depth distribution function}. \cite{Jablonski2014} The probability that a photoelectron originating at depth $z$ escapes the sample at emission angle $\vartheta$ (Fig.\,1B) is given by (!\,\excep{DDF}) \cite{Greczynski2023,Jablonski2014} 
\begin{equation}
\phi\left(z,\vartheta\right)=e^{\left(\frac{-z}{\lambda_\mathrm{i} \mathrm{cos}\,\vartheta}\right)},    
\end{equation}
that is, the signal decays exponentially with depth (Fig.\,1D), with a characteristic length determined by the inelastic mean free path $\lambda_\mathrm{i}$. \cite{Powell2020} $\lambda_\mathrm{i}$ depends on the material and the kinetic energy of the emitted electron $E_\mathrm{kin}^\ast$. 

\textbf{Background}. The longer the path that electrons travel, the greater the extent of inelastic scattering. Inelastically scattered electrons contribute to the background on the higher BE side of the peak. Backgrounds, therefore, carry information about the depth distribution of elements in the sample. \cite{Tougaard2021} For peak fitting and quantification, the background must be subtracted. This is typically done by first removing the pre-peak background \add{--- typically a constant or decreasing contribution to the background intensity originating from the peaks at lower binding energies ---} using a straight-line fit (Fig.\,1L), followed by applying an appropriate background function, such as the Shirley or Tougaard function (Fig. 1L). \cite{Tougaard2021, Engelhard2020} 

\textbf{Elastic Scattering.} Electrons can also be scattered elastically, meaning that their direction of propagation changes while their energy remains unchanged; this affects the quantification and determination of layer thickness. For emission angles $\vartheta>60\,^\circ$ elastic scattering cannot be neglected. \cite{Powell2020} To account for elastic scattering, the inelastic mean free path $\lambda_\mathrm{i}$ is replaced by the effective attenuation length (EAL) $\lambda_\mathrm{A}$. \cite{Powell2020}

\section{Quantification}

\noindent \textbf{Homogeneous samples.} For a calibrated instrument (!\,\excep{calibtrf}) and a given density of incident X-rays $I_\mathrm{X\text{-}ray}$, the photoelectron intensity of the peak (peak area) of a core level CL of an element A measured at the emission angle $\vartheta$ is given by \cite{Briggs1990,Chambers2024}
\begin{equation}
I_\mathrm{CL} = I_\mathrm{X\text{-}ray}\sigma_\mathrm{CL}L_\mathrm{CL}
\int\limits_{z=0}^{\infty}{n_\mathrm{A}\left(z\right)}\ \phi\left(z,\vartheta\right)\ \mathrm{d}z.    
\end{equation}
This intensity is proportional to the concentration depth distribution $n_\mathrm{A}\left(z\right)$ of element A in the sample and to the photoemission cross section $\sigma_\mathrm{CL}$ of CL of A; the angular photoemission asymmetry term $L_\mathrm{CL}$ is the same for all elements and \add{energy levels} if the spectrometer is in the “magic angle geometry” (!\,\excep{magicA}). Assuming a homogeneous distribution of all elements in the sample $n_\mathrm{A}\left(z\right)=const$, and removing the terms that are the same for all elements, we obtain a set of relative intensities $I_\mathrm{CL}^\mathrm{homogeneous}=n_\mathrm{A}\sigma_\mathrm{CL}\lambda_\mathrm{i}$, which are subsequently rearranged and normalized over all elements to yield the relative atomic concentrations \cite{Greczynski2023}
\begin{equation}
n_\mathrm{A}^\mathrm{rel} = \frac{\frac{I_\mathrm{CL}}{\sigma_\mathrm{CL}\lambda_\mathrm{i}}}{\sum_\mathrm{all\ elements}\frac{I_\mathrm{CL}}{\sigma_\mathrm{CL}\lambda_\mathrm{i}}}.    
\end{equation}
The term $\sigma_\mathrm{CL}\lambda_\mathrm{i}$ is often referred to as the sensitivity factor. While the process is straightforward, several critical considerations are necessary to obtain accurate information; \cite{Shard2020} typically, the absolute precision is 4–-15\,\%, \cite{Brundle2020} but the changes as small as 1\,\% can be detected between samples. \cite{Shard2020} Equation (5) is provided by most software tools, but the validity of the crucial assumption of a homogeneous sample remains the responsibility of the researchers. 

\textbf{Inhomogeneous samples.} Equation (4) can be used to derive the formulas for determining layer thickness. \cite{Powell2020} Considerable effort has been devoted to developing methodologies for determining the depth distribution of elements $n_\mathrm{A}\left(z\right)$ in the sample from a series of spectra taken at distinct angles, \cite{Herrera-Gomez2023} but this has never resulted in a universal tool. 

When XPS is combined with an ion beam, a depth profile of elements can be obtained; \cite{Shard2024} however, since the ion beam alters most of the probed volume, information about the chemical environment is typically lost. \cite{Greczynski2022}

\section{Conclusions}
\noindent X-ray photoelectron spectroscopy can be used to obtain quantitative information about the elements in a sample and to determine their chemical environments. Analyses must adhere to physical principles while considering the sample chemistry, and should be consistent. First, there should be consistency across the spectrum: all peaks associated with an element should appear with relative intensities as expected from theory (proportional to sensitivity factors $\sigma_\mathrm{CL}\lambda_\mathrm{i}$). Second, there should be consistency across the chemical states: for example, if bonding between carbon and nitrogen is identified in the C\,1s spectrum with intensity $\propto\sigma_\mathrm{C\, 1s}\lambda_\mathrm{i}$, the corresponding nitrogen bonding should be present in the N\,1s spectrum with a matching intensity ($\propto\sigma_\mathrm{N\,1s}\lambda_\mathrm{i}$). Third, consistency should be maintained across the series of samples: similar approaches, such as background subtraction or peak fitting, should be applied to all the samples unless a clear justification is provided for any differences. \add{In summary, careful and consistent analysis is necessary for producing reliable, reproducible results.}

Several critical points for correct analysis and interpretation are often overlooked: \cite{Greczynski2022,Pinder2024} (i) it is not necessary to fit all spectra; (ii) peak fitting must follow strict rules; (iii) often, two or more peaks are associated with a single chemical state: p, d, and f peaks are spin-orbit-split doublets and transition metals exhibit complex peak shapes due to multiplet splitting; (iv) normal quantitative analysis assumes a homogeneous sample; and (v) incorrect practices present in the literature should not be repeated.

\begin{acknowledgments}
\noindent We thank Brno University of Technology for support through the Excellence Research Fund.
\end{acknowledgments}

\section*{Data Availability}
\noindent Data sharing is not applicable to this article as no new data were created or analyzed in this
study.

\section*{Conflict of Interest}
\noindent The authors have no conflicts to disclose.

\appendix

\section{Notes on exceptions to general rules}
\noindent \textbf{\ref{monochrom}}. A monochromatic Al K$\upalpha$ X-ray source is standard equipment in modern instruments. However, many instruments are still equipped with a non-monochromatic Mg/Al dual X-ray source. In the X-ray spectrum of such sources, a single K$\upalpha_{1,2}$ line is dominant, but weaker lines such as K$\upalpha_{3,4}$ and K$\beta$ are also present; these additional lines also excite photoelectrons, resulting in non-monochromatic satellite features that appear at the lower-BE side of the peaks excited with the main X-ray line. \cite{Stevie2020} Dual anode sources offer two distinct X-ray energies, which can be used to distinguish Auger peaks from photoelectron peaks. 
Alternatively, synchrotron radiation can be used as an X-ray source, offering superior energy resolution and enabling advanced techniques beyond standard XPS. When using polarized X-rays with tunable energy, several parameters change: for example, photoionization cross sections become energy dependent, and both the angular asymmetry parameter \cite{Shard2020} and effective attenuation lengths (EALs) \cite{Powell2020} differ from those for unpolarized laboratory sources.\\[1mm]
\noindent \textbf{\ref{charneut}}. The exit part of the dual anode X-ray source, which consists of an aluminum window positioned close to the sample, generates a significant number of stray electrons. \cite{Stevie2020} When sample charging occurs, these electrons are attracted to the sample and partially neutralize it. As a result, the measured spectrum is shifted by up to $\sim10$\,eV towards higher BEs.\\[1mm]
\noindent \textbf{\ref{neutral}}. \add{In the majority of cases, the charge neutralization does not result in a neutral surface; hence, the spectrum is shifted from the nominal position. 
By differential charging, we mean a non-homogeneous distribution of charge within the analyzed area, i.e., that some areas are charged differently than others. The reasons for that may be instrumental, i.e., non-uniform X-ray flux over the analyzed area, or sample-related, i.e., non-homogeneous samples.} \\[1mm]
\noindent \textbf{\ref{photoem}}. Photoemission is a quantum-mechanical process, which sometimes results in slightly different separations and intensity ratios than those predicted by the state multiplicity rule. \cite{Pinder2024} An example can be the broadening of Ti\,2p$_{1/2}$. \cite{Chambers2025} 
It should be noted that although the higher BE component of the doublet may experience slight broadening, \cite{Major2020a} the intensity ratio (given by the peak areas) remains that predicted by the state multiplicity rule. \\[1mm]
\noindent \textbf{\ref{chemshift}}. Because photoemission is an inherently quantum mechanical effect, simplified models may not account for all observed phenomena. \cite{Bagus2023} \\[1mm]
\noindent \textbf{\ref{Ovac}}. \add{A common misconception is the direct assignment of a particular peak in the O 1s spectrum to oxygen vacancies. Oxygen vacancies are not directly detectable in the O\,1s spectrum, as there is no photoemission from a missing atom, and next-nearest-neighbor effects are also not present. \cite{Idriss2021,Easton2025}}\\[1mm]
\noindent \textbf{\ref{C1senergy}}. \add{With the reported standard deviation of $\sigma=0.25$\,eV, \cite{Biesinger2022} the prediction interval for a single measured value at the 95\,\% probability level is (284.9$\pm$0.5)\,eV, considering also the 0.1\,eV accuracy in the absolute value of measured energy\cite{Seah2001} and normal distribution. The 99.7\% prediction interval would be $\pm$0.8\,eV. Note that other authors report even larger standard deviations for the C 1s peak position of AdC. \cite{Biesinger2022}}\\[1mm] 
\noindent \textbf{\ref{theosensf}}. When using theoretical sensitivity factors. \cite{Brundle2020} See point \ref{calibtrf}. \\[1mm]
\noindent \textbf{\ref{DDF}}. Depth distribution function in its simplest form – a straight-line approximation neglecting elastic scattering. \cite{Jablonski2014} \\[1mm]
\noindent \textbf{\ref{calibtrf}}. \add{A properly set-up XPS instrument includes calibration of the energy scale, \cite{Stevie2020} detector linearity, \cite{Shard2020,Greczynski2022} and the transmission function.} For the transmission function, some instruments are calibrated against a “true spectrum” (a reference spectrum provided by NIST); in this case, theoretical sensitivity factors can be used for quantification. \cite{Shard2020} However, some instruments are calibrated to the manufacturer’s reference; in this case, instrument-bound sensitivity factors should be used. 
\add{Instruments from different manufacturers can differ in how the transmission function is handled and applied: either to directly correct the spectra or to correct the sensitivity factors.
Therefore, it is important to learn what the instrument does with data and what kind of data is exported.}  \\[1mm]
\noindent \textbf{\ref{magicA}}. If the angle between the X-ray source and spectrometer is 54.7\,$^\circ$, the angular asymmetry is the same for all lines and elements. \cite{Briggs1990} \\[1mm]
\noindent \textbf{23}. This is an example of a reference given in the introduction.

\bibliography{refs}

\end{document}